\documentclass[pdflatex,sn-mathphys]{sn-jnl}% Math and Physical Sciences Reference Style
\usepackage{algorithm}
\usepackage{algorithmicx}

\usepackage{amsmath,amssymb,amsfonts}
\usepackage{graphicx}
\usepackage{textcomp}
\usepackage{xcolor}

\newcommand{\stitle}[1]{\vspace{1ex} \noindent{\bf #1}}

\usepackage{booktabs} % For formal tables
\usepackage{bbm}
\usepackage{epstopdf}
\usepackage{url}
\usepackage{multirow}
\usepackage{subfigure}

\jyear{2022}%

\theoremstyle{thmstyleone}%
\theoremstyle{thmstyletwo}%
\newtheorem{example}{Example}%

\theoremstyle{thmstylethree}%
\newtheorem{definition}{Definition}%

\begin{document}

\title[SECS on Temporal Networks: Concepts and Algorithms]{Significant Engagement Community Search on Temporal Networks: Concepts and Algorithms}


%\author*[1]{\fnm{Pingpeng} \sur{Yuan}}\email{ppyuan@hust.edu.cn}

\author[1]{\fnm{Yifei} \sur{Zhang}}\email{yfzhangsz@hust.edu.cn}
\equalcont{These authors contributed equally to this work.}
\author[1]{\fnm{Longlong} \sur{Lin}}\email{longlonglin@hust.edu.cn}
\equalcont{These authors contributed equally to this work.}

\author*[1]{\fnm{Pingpeng} \sur{Yuan}}\email{ppyuan@hust.edu.cn}

\author[1]{\fnm{Hai} \sur{Jin}}\email{hjin@hust.edu.cn}

\affil[1]{\orgdiv{National Engineering Research Center for Big Data Technology and System,
Services Computing Technology and System Lab,
Cluster and Grid Computing Lab, School of Computer Science \& Technology}, \orgname{HuaZhong University of Science and Technology}, \orgaddress{\street{Luoyu Road 1037}, \city{Wuhan}, \postcode{430074}, \state{Hubei}, \country{China}}}

\abstract{Community search, retrieving the cohesive subgraph which contains the query vertex, has been widely touched over the past decades. The existing studies on community search mainly focus on static networks. However, real-world networks usually are temporal networks where each edge is associated with timestamps. The previous methods do not work when handling temporal networks. We study the problem of identifying the significant engagement community to which the user-specified query belongs. Specifically, given an integer $k$ and a query vertex $u$, then we search for the subgraph $\mathcal{H}$ which satisfies (i) $u \in \mathcal{H}$; (ii) the de-temporal graph of $\mathcal{H}$ is a connected $k$-core; (iii) In $\mathcal{H}$ that $u$ has the maximum engagement level. To address our problem, we first develop a top-down greedy peeling algorithm named \textit{TDGP}, which iteratively removes the vertices with the maximum temporal degree. To boost the efficiency, we then design a bottom-up local search algorithm named \textit{BULS} and its enhanced versions \textit{BULS+} and \textit{BULS*}. Lastly, we empirically show the superiority of our proposed solutions on six real-world temporal graphs.}

\keywords{Graph mining, Temporal networks, Community search, $k$-core}

\maketitle

\section{Introduction}
\label{sec:intro}
There are numerous community structures (i.e., densely connected subgraphs) presented in real-world networks. Therefore, mining communities is an
important tool for analyzing network structure and organization. Generally,
there are two main  research directions on community mining: (1) community
detection  identifies all communities by some predefined criteria \cite{newman2004fast, donetti2004detecting, rokach2005clustering, DBLP:conf/icde/ChangQ19}. However, it has intractable computational bottleneck and is
not customized for user-specified query vertices. (2) community search  aims to identify the community containing the
user-specified query vertices \cite{DBLP:conf/kdd/SozioG10, DBLP:conf/sigmod/CuiXWW14}, which is more efficient and personalized. Besides, community search can also be applied to numerous high-impact applications, including friend recommendation, link-spam detection, and drug discovery.

Despite the significant success of community search,
all of these approaches are performed under the context of static graphs. However, the relationships of real-world networks vary over time. For instance, a researcher collaborates with others on a project or a paper at some time. Persons call their friends from time to time. Such time-related connections among entities can be naturally modeled as temporal graphs \cite{DBLP:conf/kdd/RozenshteinG19,  9424972, DBLP:conf/icde/LiSQYD18,DBLP:conf/kdd/YangYWCZL16}, in which each edge is attached a timestamp to indicate when  the connections occur.
In such networks, an entity actively engages in a community via frequent connections with other entities at different periods while others may incur occasional relationships. Moreover, the entity has different engagement levels in different communities. And the engagement level of the entity  will affect whether other entities are also engage in the community. For example, on social networks (e.g. Facebook, Microblog), community members frequently post timely and interesting information, and the community clearly become more active and appealing. When few community members share interesting and latest contents, persons cannot find the information that appeals to them. Then they may leave the community. 
As a result, it is more useful and challenging to study the engagement level of the entity in a community and identify the target community with the highest engagement level from all communities.

To this end, we introduce a new problem of identifying the significant engagement community to which the user-specified query belongs. Specifically,  given a temporal graph $\mathcal{G}$, an integer $k$ and a query vertex $u$, the significant engagement community of $u$ is a temporal subgraph $\mathcal{H}$ which satisfies (i) $u \in \mathcal{H}$; (ii) the de-temporal graph of $\mathcal{H}$ is a connected $k$-core ($k$-core is the subgraph in which each vertex has a degree not smaller than $k$ \cite{DBLP:journals/corr/cs-DS-0310049}); (iii) In $H$ that $u$ has the maximum temporal engagement (more details in Section \ref{sec:eng}). Clearly, condition (i) requires $\mathcal{H}$ is the community where the query vertex $u$ is located. Condition (ii) captures the cohesiveness of the community by the representative cohesive subgraph $k$-core \cite{DBLP:journals/corr/cs-DS-0310049}, which is reasonable. Condition (iii) is our objective function, which requires the result to be the community with the highest engagement level for the query vertex $u$. 

Surprisingly, detecting significant engagement community of the query vertex is
meaningful and enjoys many applications, but this issue has
not been adequately studied in literature. As stated in Section \ref{sec:relate}, the existing approaches on community search \cite{DBLP:conf/icde/HuangLX17, DBLP:journals/vldb/FangHQZZCL20} only consider the structural cohesiveness
but the temporal feature of a subgraph. Until very recently,
some studies were done on identifying temporal community. For example,  Qin et al. \cite{9210070} raised periodic communities which regular arise periodically, this can be used to handle periodic phenomenons. Li et al. \cite{DBLP:conf/icde/LiSQYD18} raised the persistent community model which the communities are stable over time. Chu et al. \cite{DBLP:journals/pvldb/ChuZYWP19} researched the problem of detecting density bursting subgraphs, where the density bursting subgraphs are defined as those which accumulate their density at the fastest speed. Unfortunately, all these approaches lack the attention to the relationships between the vertices inside and the subgraphs, they cannot tell the differences how the vertices participate in those subgraphs. Thus, it is unclear how to adopt the existing technologies to solve our problem. In a nutshell, our contributions are
reported as follows:

\stitle{\underline{Novel model.}} 
We propose a novel community search model named  \textit{SECS}, which comprehensively considers the structural cohesiveness of the community and the engagement level of the query vertex.  \\
\stitle{\underline{Effective algorithms.}} 
We first propose a top-down greedy peeling algorithm  \textit{TDGP} and then design a more efficient bottom-up local search algorithm \textit{BULS}, and its enhanced versions \textit{BULS+} and \textit{BULS*} with some powerful expanding strategies. \\
\stitle{\underline{Extensive experiments.}} 
We conduct extensive experiments on six real-world temporal graphs, which reveal that our solutions perform well in terms of both the efficiency and effectiveness. 

This paper is an expanded version of the conference paper \cite{DBLP:conf/dasfaa/ZhangLYJ22} written by Zhang et al. . Compared with the conference paper, in this paper we proposed a new local search algorithm named \textit{BULS*}, which uses a new advanced expanding strategy called occurrence-driven strategy. As for the experiment content, we add the scalability testing in both the efficiency testing and effectiveness testing. And in effectiveness testing, we involved a new metric called temporal conductance \cite{DBLP:conf/www/SilvaSS18} to evaluate the quality of the results from a different perspective. We also add the comparison of the effectiveness of different temporal methods with default parameters, the comparison of the size of expanded graph, and the case study on DBLP to better demonstrate the value of our research. Besides, there are more results and analysis in each experiment. The experiments show that the new algorithm \textit{BULS*} can get the best results with the least running time. In addition, we have a more substantial and detailed introduction in other part of the article.

\section{Significant Engagement Community Search} \label{sec:eng}

\subsection{Preliminaries}
Here, we consider an undirected temporal graph $\mathcal{G}(\mathcal{V},\mathcal{E},\mathcal{T})$, in which $\mathcal{V}$ is the set of vertices inside the $\mathcal{G}$, 
$\mathcal{E}=\{(u,v,t)\mid u,v\in \mathcal{V}\}$ is the set of temporal edges and $\mathcal{T}=\{t\mid (u,v,t)\in \mathcal{E}\}$ is the timestamps set of $\mathcal{G}$.
Additionally, we define $\mathcal{H}=(\mathcal{V_H},\mathcal{E_H}, \mathcal{T_H})$ as the temporal subgraph of $\mathcal{G}$ when $\mathcal{V_H}\subseteq \mathcal{V}$, 
$\mathcal{E_H}\subseteq \mathcal{E}$, and $\mathcal{T_H}\subseteq \mathcal{T}$. $\mathcal{G}$'s de-temporal graph is $G(V,E)$, which meets the conditions that: $V=\mathcal{V}$ and $E=\{(u,v)\mid \exists (u,v,t)\in \mathcal{E}\}$. Namely, the de-temporal graph $G$ is a static graph that ignores the temporal information carried on the edges. Similarly, we denote $H(V_H,E_H)$ as a subgraph of $G$ when it satisfies $V_H \subseteq V$ and $E_H \subseteq E$. To help formalize our problem, we put out several definitions as follows.   
\iffalse
\begin{figure}[t]
  \centering 
  \subfigure[Temporal graph $\mathcal{G}$]{ 
    \includegraphics[scale=0.2]{{fig-ex-1.png}} 
    \label{fig:example:a}
  }
  \hspace{0.1cm}
  \subfigure[Cumulative graph $C^{3}_{0}$]{ 
    \includegraphics[scale=0.2]{{fig-ex-2.png}} 
    \label{fig:example:b}
  } 
  \hspace{0.1cm}
  \subfigure[Cumulative graph $C^{4}_{2}$]{ 
    \includegraphics[scale=0.2]{{fig-ex-3.png}} 
    \label{fig:example:c}
  } 
  \caption{Example of temporal graph $\mathcal{G}$ and its cumulative graphs} 
  \label{fig:s_fig2} 
  \vspace{-0.6cm}
\end{figure}
\fi
\iffalse
\begin{figure}[h]
\centering
\includegraphics[scale=0.2]{fig-ex-1.png}
\caption{Temporal graph $\mathcal{G}$}\label{fig:example:a}
\end{figure}

\begin{figure}[h]
\centering
\includegraphics[scale=0.2]{fig-ex-2.png}
\caption{Cumulative graph $C^{3}_{0}$}\label{fig:example:b}
\end{figure}

\begin{figure}[h]
\centering
\includegraphics[scale=0.2]{fig-ex-3.png}
\caption{Cumulative graph $C^{4}_{2}$}\label{fig:example:c}
\end{figure}
\fi

\begin{definition}[Edge Occurrences]
\label{def:ef}
 Edge occurrence is a measure to demonstrate how many times the connections between two vertices occur within an interval of time. We first define the following function to indicate whether an edge exists:
 \begin{equation}
    \pi(u,v,t)=\left \{
   \begin{array}{cl}
   0& ~~~(u,v,t) \notin \mathcal{E}\\
   1&~~~(u,v,t) \in \mathcal{E} \\  \end{array} \right.
  \end{equation}

So, the edge occurrences of $(u,v)$ over time interval $[t_s,t_e]$ is defined as following:
\begin{equation}
    o_{(u,v)}(t_s,t_e)=\sum_{i=t_s}^{t_e} \pi(u,v,i)
\end{equation}
\end{definition}

\begin{definition}[Cumulative Graph]
\label{def:cg}The cumulative graph of temporal graph $\mathcal{G}$ for time interval $[t_s,t_e]$ is a weighted graph  $C_{t_s}^{t_e}(\mathbb{V}_{t_s}^{t_e},\mathbb{E}_{t_s}^{t_e},w_{t_s}^{t_e})$, in which the $\mathbb{V}_{t_s}^{t_e}=\{ u\mid(u,v,t)\in \mathcal{E},t\in [t_s,t_e] \}$, $\mathbb{E}_{t_s}^{t_e}=\{ (u,v)\mid(u,v,t)\in \mathcal{E},t\in [t_s,t_e] \}$, and  $w_{t_s}^{t_e}(u,v) = o_{(u,v)}(t_s,t_e)$. Let $C_{\mathcal{G}}$ be the cumulative graph of $\mathcal{G}$ when the time interval is [min($\mathcal{T}$),max($\mathcal{T}$)]. Additionally, we have $\mathbb{N}_{u,C_{\mathcal{H}}}=\{v\mid(u,v,t) \in \mathcal{E_H}\}$ and $\mathbb{D}_{u,C_{\mathcal{H}}}=\mid\{(u,v)\mid(u,v,t) \in \mathcal{E_H}\}\mid$. 
\end{definition}

\begin{definition}[Temporal Degree]
\label{def:td}

The temporal degree of the vertex $u$ w.r.t. $[t_s,t_e]$ and temporal graph $\mathcal{G}$ is defined as following:
\begin{equation}
\label{eq:td}
    d_{u,\mathcal{G}}(t_s,t_e)=\sum_{i=t_s}^{t_e} \pi(u,v,i)=\sum_{v \in \mathcal{G}}w_{t_s}^{t_e}(u,v)
\end{equation}

So, temporal degree of $u$ in $\mathcal{G}$ is $d_{u,\mathcal{G}}=d_{u,\mathcal{G}}(min(\mathcal{T_G}),max(\mathcal{T_G}))$.
 
\end{definition}

\begin{figure}[!t]
    \begin{minipage}{0.32\linewidth}
    \centerline{\includegraphics[width=4.0cm]{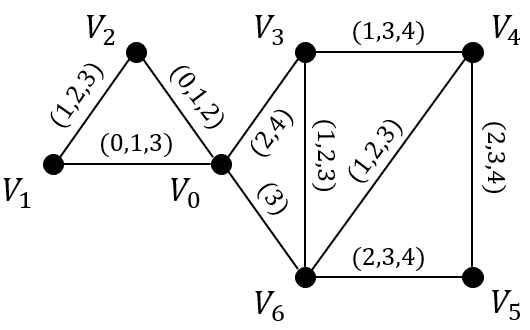}}
    \centering
    \caption{Temporal graph $\mathcal{G}$}\label{fig:example:a}
    \end{minipage}
    \hfill
    \begin{minipage}{0.32\linewidth}
    \centerline{\includegraphics[width=4.0cm]{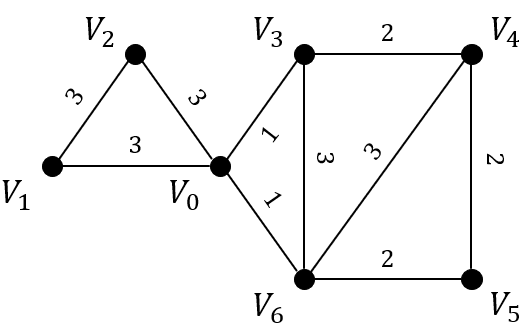}}
    \caption{Cumulative graph $C^{3}_{0}$}\label{fig:example:b}
    \end{minipage}
    \hfill
    \begin{minipage}{0.32\linewidth}
    \centerline{\includegraphics[width=4.0cm]{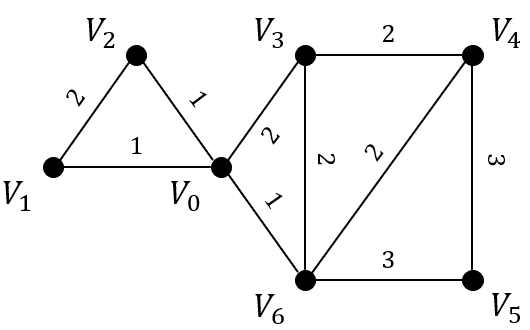}}
    \caption{Cumulative graph $C^{4}_{2}$}\label{fig:example:c}
    \end{minipage}
\end{figure}

\begin{example}
 Fig. \ref{fig:example:a} shows a temporal graph $\mathcal{G}$ in which there are 7 vertices with 27 temporal edges, Fig. \ref{fig:example:b} and Fig. \ref{fig:example:c} are the cumulative graphs of $\mathcal{G}$ with time interval [0, 3] and [2, 4] respectively. Let's take a look at the $\mathcal{G}$ in Fig. \ref{fig:example:a}, there are three temporal edges linking $V_{0}$ and $V_{1}$: $(V_{0},V_{1},0)$,$(V_{0},V_{1},1)$, and $(V_{0},V_{1},3)$. Thus, the edge occurrences of $(V_0,V_1)$ over $[0,4]$ is $o_{(0,1)}(0,4)=3$. For the vertex $V_0$, its temporal degree  $d_{V_0,\mathcal{G}}=9$. 
\end{example} 

\subsection{Problem definition}

The vertices of a community may participate in the community in different ways. The more engaged they are in the community, the more active they will help to shape the community to be more active. Generally, their community engagement varies in different communities. So, we introduce engagement level to evaluate impact on a community of a vertex.
\begin{definition}[Engagement Level]
\label{def:eng} For the temporal subgraph $\mathcal{H}$,
engagement level of vertex $u$ in $\mathcal{H}$ is the impact on $\mathcal{H}$ which $u$ achieves. It is defined as:
\begin{equation}
    Eng_u(\mathcal{H})=\frac{d_{u,\mathcal{H}}}{\sum_{v\in \mathcal{H}}d_{v,\mathcal{H}}}
\end{equation}
\end{definition}

The engagement level $Eng_u(\mathcal{H})$ only indicates the activity of $u$ in $\mathcal{H}$. However, high engagement level does not mean that $\mathcal{H}$ is a cohesive subgraph. Thus, we adopt one representative model $k$-core \cite{DBLP:journals/corr/cs-DS-0310049} to model the cohesiveness of the community. By doing this, we proposed a more practical model by considering both structural cohesiveness and temporal engagement level.
\begin{definition}[$k$-core \cite{DBLP:journals/corr/cs-DS-0310049}] 
\label{def:kcore}
For a de-temporal graph $G$, $H$ is a subgraph of $G$. We say $H$ is a $k$-core in $G$ if $\mid\{v\mid(u,v)\in H\}\mid\geq k$ for any vertex $u \in H$ holds.

\end{definition}

\begin{example}
Considering the temporal graph $\mathcal{G}$ in Fig. \ref{fig:example:a}, we have a temporal subgraph $\mathcal{H}$ that $\mathcal{V_H}=\{ V_0,V_1,V_2\}$, $\mathcal{T_H}=\mathcal{T_G}$, and $\mathcal{E_H}=\{(u,v,t) \mid (u,v,t)\in \mathcal{E_G}, u,v \in \mathcal{V_H}, t\in \mathcal{T_H} \}$, we have $Eng_{V_0}(\mathcal{H})=\frac{6}{18}=\frac{1}{3}$. Meanwhile, we can observe that $\mathcal{H}$ is a $2$-core;
\end{example}

Based on Definition \ref{def:eng} and \ref{def:kcore}, we formulate our problem as follows. 

\stitle{\textbf{Our problem (Significant Engagement Community Search: \textit{SECS}).}} Given a temporal graph $\mathcal{G}$, a query vertex $u$, and a parameter $k$, our goal is to find a temporal subgraph $\mathcal{H}$ which meets: 
i) $u \in \mathcal{V_H}$; ii) the de-temporal graph of $\mathcal{H}$ is a connected $k$-core;
iii) $Eng_u(\mathcal{H})\geq Eng_u(\mathcal{H'})$ for all temporal subgraph $\mathcal{H}'$. For simplicity, we call $\mathcal{H}$ is a  significant engagement community ($\mathcal{SEC}$ for short) of $u$.

To solve this problem, we need to evaluate different temporal subgraphs with different time periods and node sets.

\section{The Top-Down Greedy Peeling Algorithm}

\subsection{Overview}
In this subsection at first, we introduce the overall framework of our proposed top-down greedy peeling algorithm (\textit{TDGP}), which is shown in Algorithm \ref{al:frame}. In order to get the significant engagement community ($\mathcal{SEC}$), the first thing we need to do is generating the cumulative graphs from the temporal graph $\mathcal{G}$. Since there are $\mid\mathcal{T}\mid$ timestamps, we can get in total $(1+\mid\mathcal{T}\mid)\mid\mathcal{T}\mid/2$ time intervals. Each time interval corresponds to a cumulative graph. Considering the cohesive constraint for $\mathcal{SEC}$, for each cumulative graph $C_{\mathcal{H}}$, we need to maintain the de-temporal graph $\mathcal{H}$ as a $k$-core and check whether it contains the query vertex $u$. When $C_{\mathcal{H}}$ meets all these requirements, it comes into the next stage to reduce its extent to maximize $Eng_u(\mathcal{H})$. For this part, since there is no direct correlation between $d_{u,\mathcal{H}}$ and $Eng_u(\mathcal{H})$, we delete the vertices with the maximum temporal degree greedily in order to maximize $d_{u,\mathcal{H}}$, until it can not satisfy the conditions for $\mathcal{SEC}$ mentioned above. After all of these, we can finally get the community $\mathcal{SEC}$ in which $u$ has the maximum engagement level.

\begin{algorithm}[t!]
\caption{Top-Down Greedy Peeling Algorithm}
\label{al:frame}
\begin{algorithmic}[1]
\Require temporal graph $\mathcal{G}$, query vertex $u$, integer $k$
\Ensure significant engagement community $\mathcal{SEC}$
\State $\mathcal{C} \leftarrow $compute all the cumulative subgraphs of $\mathcal{G}$
\For{each $C_{\mathcal{H}}\in\mathcal{C}$ contains $u$}
\While{$C_{\mathcal{H}}$ is a $k$-core contains $u$} 
\State select a vertex $v$ ($v \neq u$) with the maximum temporal degree
\State $C_{\mathcal{H}} \leftarrow C_{\mathcal{H}}-v$
\EndWhile
\EndFor
\State $\mathcal{SEC} \leftarrow $ $arg max_{C_{\mathcal{H}}}Eng_u(\mathcal{H})$
\State return $\mathcal{SEC}$
\end{algorithmic}
\end{algorithm}

We only consider the time intervals of which $u$ has edges occur on their two ends. Though we cannot make sure that with this pruning strategy whether some $k$-core structures are ruined, however in this way we can pay attention to the time interval that $u$ has action instead of the whole time interval of $\mathcal{G}$. Which will markedly reduce the useless work on dealing other time intervals, the cost that some possible results may be pruned is insignificant comparing to the benefits.

As for the process of computing the cumulative graphs in Algorithm \ref{al:frame}, an efficient way to handle this is to calculate a new cumulative graph based on an extant one, if not, we will consume much time to generate the new cumulative graphs. In detail, within $\mathcal{G}$, from the cumulative graph $C^{t_b}_{t_a}$, we can easily get the cumulative graph $C^{t_b-1}_{t_a}$ and $C^{t_b}_{t_a+1}$ which only need to let the weight of edges minus one if there exists temporal edges occur in the time $t_b$ or $t_a$. That is, when we start to calculate the cumulative graph at the beginning of our algorithm, we get the first cumulative graph $C_{\mathcal{G}}$. After that we can respectively calculate another two cumulative graphs $C^{max(\mathcal{T})-1}_{min(\mathcal{T})}$ and $C^{max(\mathcal{T})}_{min(\mathcal{T})+1}$. Specifically, take the process of dealing $C^{max(\mathcal{T})-1}_{min(\mathcal{T})}$ for example, at first we let $\mathbb{V}^{max(\mathcal{T})-1}_{min(\mathcal{T})}=\mathbb{V}^{max(\mathcal{T})}_{min(\mathcal{T})}$ and $\mathbb{E}^{max(\mathcal{T})-1}_{min(\mathcal{T})}=\mathbb{E}^{max(\mathcal{T})}_{min(\mathcal{T})}$, for the edges $(u,v) \in \mathbb{E}^{max(\mathcal{T})-1}_{min(\mathcal{T})}$.
\begin{equation}
    w^{max(\mathcal{T})-1}_{min(\mathcal{T})}(u,v)=\left \{
   \begin{array}{cl}
    w^{max(\mathcal{T})}_{min(\mathcal{T})}(u,v)& ~~~(u,v,b) \notin \mathcal{E}\\
   w^{max(\mathcal{T})}_{min(\mathcal{T})}(u,v)-1&~~~(u,v,b) \in \mathcal{E} \\  \end{array} \right.
  \end{equation}
After these, we can remove the edges in $\mathbb{E}^{max(\mathcal{T})-1}_{min(\mathcal{T})}$ whose weights are zero and then remove the vertices with no edge linking them in $\mathbb{V}^{max(\mathcal{T})-1}_{min(\mathcal{T})}$. 

\subsection{Analysis}
Here we take $m$ to represent $\mid\mathcal{T}\mid$ and $n$ to represent the scale of the graph. In this algorithm, we need to deal with in total $(1+\mid\mathcal{T}\mid)\mid\mathcal{T}\mid/2$ amount of cumulative graphs. For each one of them, we greedy delete the vertices with the maximum temporal degree, so the time complexity for the whole algorithm is $O(nm^{2})$. We use a mitosis and BFS way to consider the time interval, there are at most $1+\mathcal{T}$ times of accumulate graphs exists at one time, the whole space complexity of the $\textit{TDGP}$ is $O(nm)$.

\begin{algorithm}[t!]
\caption{Naive Candidate Generation Algorithm}
\label{alg: naive candidate}
\begin{algorithmic}[1]
\Require cumulative graph $C_{\mathcal{H}}$, query vertex $u$, integer $k$
\Ensure alternative subgraph $\mathcal{AS}$

\State $AS \leftarrow \emptyset$;$Q\leftarrow \emptyset$
\State $Q$.push($u$); $\mathcal{AS}$.push($u$)
\While{$Q \neq \emptyset$}
\State $s \leftarrow Q$.pop()
\For{each $v \in$ $\mathbb{N}_{s,C_\mathcal{H}}$}
\If{$\mathbb{D}_{v,C_\mathcal{H}} \geq k$}
\State $Q$.push($v$); $\mathcal{AS}$.push($v$)
\EndIf
\EndFor
\EndWhile
\State return $\mathcal{AS}$
\end{algorithmic}
\end{algorithm}

\section{The Bottom-Up Local Search Algorithm}

\subsection{Overview}

In this subsection, we develop a bottom-up local search algorithm (\textit{BULS}) to reduce the scale of our cumulative graph $C_{\mathcal{H}}$ in the greedy algorithm, before $C_{\mathcal{H}}$ being iteratively removed vertices from. The core concept of local search method is to generate an alternative subgraph $\mathcal{AS}$ from the query vertex $u$, then the same steps in Algorithm \ref{al:frame} will be proceed to receive our results. It should be noticed that since we use the approximate strategy to delete the redundant vertices in both the global and local algorithms, there might be some differences between the final results using the two methods to get respectively in the same cumulative graph.   

Since that the de-temporal graph of maximal engagement community is a $k$-core, in which each vertex has the degree no less than $k$. 
Obviously, the vertices with degrees less than $k$ in the de-temporal graph of $\mathcal{G}$ can never be included in our result, so they can be straightly pruned. When further expanding the analysis on our problem, considering the temporal attributes and the inclusion relation among the cumulative graphs, that their degrees might be lower in the cumulative graphs with shorter time interval. For the process of dealing each cumulative graphs, we need to check that whether exists for a vertex $v$ that $\mathbb{D}_{v,C_{\mathcal{H}}}<k$, if so we need to prune it. 
Here a simple way to generate the alternative graph has come out, that only chooses the vertices with degrees no less than $k$ to be included into our alternative subgraph. The specific procedures are shown in Algorithm \ref{alg: naive candidate}, which is called naive candidate generation algorithm for it only use the degree criterion to decide whether to choose the vertices as candidates. In detail, we begin expanding the alternative graph from the query vertex $u$ and use a BFS way to traverse other vertices.

It's clear that this algorithm lacks efficiency due to only considering the structural properties. Therefore, we develop the candidate generation algorithm additionally using the relationships among the temporal degrees and vertex engagement. According to the fraction of engagement level in Definition \ref{def:eng}, we want the temporal degree of query vertex $u$ be uninfluenced in the process of pruning. So we put all the vertices connect to $u$ and with the degree no smaller than the parameter $k$ in $C_{\mathcal{H}}$ into our alternative subgraph $\mathcal{AS}$. Since the value of temporal degree is made up of the occurrences of temporal edges connecting to it, so when we put all the related vertices into the alternative subgraph, we can know the upper bound for $d_{u,\mathcal{H}}$, that $d_{u,\mathcal{H}} \leq d_{u,\mathcal{AS}}$. With it we can carry on following steps.

\begin{algorithm}[t!]
\caption{Advanced Candidate Generation Algorithm}
\label{alg:advanced}
\begin{algorithmic}[1]
\Require cumulative graph $C_{\mathcal{H}}$, query vertex $u$, integer $k$
\Ensure alternative subgraph $\mathcal{AS}$

\State $\mathcal{AS} \leftarrow \emptyset$;$S \leftarrow \emptyset$;$K \leftarrow \emptyset$;$Q \leftarrow \emptyset$
\For{each $x \in$ $\mathbb{N}_{u,C_{\mathcal{H}}}$}
\If{$\mathbb{D}_{x,C_\mathcal{H}} \geq k$}
\State $N$.push($x$)
\EndIf
\EndFor
\For{each $s \in N$}
\If{$\mathbb{D}_{s,N} \geq k$}
\State $AS$.push($s$)
\EndIf
\If{$\mathbb{D}_{s,N} < k$}
\State $K$.push($s$)
\EndIf
\EndFor
\For{each $x \in K$}
\State $\textbf{Expanding}(x)$
\EndFor
\State return $\mathcal{AS}$
\\ ~~
\State Procedure $\textbf{Expanding}(x)$
\State $Q$ $\leftarrow \emptyset$
\State $Q$.push($x$);$\mathcal{AS}$.push($x$)
\While{$Q \neq \emptyset$}
\State $m \leftarrow Q$.pop()
\For{$n \in \mathbb{N}_{m,C_\mathcal{H}}$}
\State Using expanding strategies on $n$
\EndFor
\EndWhile
\end{algorithmic}
\end{algorithm}

 We have a property for further expanding. For the cumulative graph $C_{\mathcal{H}}$,there is a subgraph $\mathcal{N}$ composed of the vertex $u$ and $\mathbb{N}_{u,C_{\mathcal{H}}}$, for the vertex $m \in \mathcal{N}$ with $\mathbb{D}_{m,\mathcal{N}}\geq k$ it lacks efficiency to expand the alternative graph proactively based on it. The reason for this is that since the vertices $m$ has already been in a $k$-core structure, their rest main contribution are the edge occurrences of which link to the query vertex $u$. For the edges which are not directly connect to the query vertex $u$, their value might appear in $\sum_{u\in \mathcal{H}}d_{u,\mathcal{H}}$, that the denominator of the fraction of $Eng_u(\mathcal{H})$. Since our algorithms are greedy ones and we want the amount of these edges as small as possible, there is no need to start to consider other vertices based on them. Intuitively, we want to minimize the sum to benefit the final result. Considering to the integrity constraint of $k$-core, we need to evaluate these vertices if it's possible for them to be included in the alternative graph. It should be noticed that for the vertex $v$ directly connecting to $u$ with $\mathbb{D}_{v,C_{\mathcal{H}}} \geq k$ and $\mathbb{D}_{v,\mathcal{N}} < k$, some of its neighbors are not connecting to $u$. That's to say without these vertices $v$ can not be involved in the alternative graph, which means we need to scale up the alternative subgraph from such vertices.
The algorithm \ref{alg:advanced} is the overview of our advance candidate generation algorithm. At the beginning, we let the query vertex $u$ and its neighbors with degree no less than $k$ in $C_{\mathcal{H}}$ to form a private community $N$. For the vertex $v\in N$, if its degree in $N$:$\mathbb{D}_{v,N}\geq k$, we do not need to further extend from them. Meanwhile, for the vertices $s$ that $s\in N$ and $\mathbb{D}_{v,N}\geq k$, we choose to start to expand from, we put them into a queue $K$. Then, at each time we pop a vertex from $K$ and use a queue $Q$ to separately handle its neighbors. With the local search process, we can eliminate the vertices which have no chance to get involved in our result according to the current judgment conditions. Specifically, we apply three expanding strategies mentioned later to determine whether to involve the new vertices into our search queue. We take $\mathcal{AS}$ to represent the temporal graph induced by the vertices in it with the temporal edges occur in [$min(\mathcal{T_H}),max(\mathcal{T_H})$]. For each vertex which satisfies our requirement, we will involve it into our search space $\mathcal{AS}$ and push it into the queue $Q$ in order to consider its neighbors in following steps. 

\subsection{Expanding strategies}

In this subsection, we introduce some effective expanding strategies in algorithm \ref{alg:advanced}.

\stitle{\textbf{Reference Strategy.}} Considering that we use a top-down way to deal different cumulative graphs for $\mathcal{H}$, there are some non-terminal results for $\mathcal{SEC}$ in this progress. We use $bestresult$ to represent the vertex engagement level for $u$ in current $\mathcal{SEC}$, thus we can use the $bestresult$ to judge whether to include more vertices. Specifically, we only consider the increment for a single vertex at each time, here is the strategy: 
for a vertex $n$ (line 26), if $n\notin \mathcal{AS}$, $d_{n,C_{\mathcal{H}}}$ $\geq$ $k$, and
    $\frac{d_{u,\mathcal{AS}}}{\sum_{v\in \mathcal{AS}}d_{v,\mathcal{AS}}+w_{C_{\mathcal{H}}}(m,n)}>bestresult$, we execute $Q$.push($n$) and $\mathcal{AS}$.push($n$).\\ 
    
    It should be noticed that the $\mathcal{AS}$ we consider is not a fixed one, and we take all the vertices in $N$ into the $\mathcal{AS}$ before run the procedure Expanding.
    We call this strategy the reference strategy for that it refers to previous non-terminal results to make decisions. 

\stitle{\textbf{Engagement-driven Strategy.}} Assume we start to expand the alternative subgraph $\mathcal{AS}$ from $x$ (line 15), when we consider a vertex $n$ (line 26), we consider the vertices in a line from $x$ to $n$ as a whole. The increment for $d_{u,\mathcal{AS}}$ is $w_{C_{\mathcal{H}}}(u,x)$ and increment for $\sum_{v\in \mathcal{AS}}d_{v,\mathcal{AS}}$ is at least the sum of edge weights one way to connect from $u$ to $n$, here we take $ac(m)$ to represent the sum from $u$ to $m$. Considering that we want the $Eng_{u}(\mathcal{AS})$ not to decrease in each step, that's to say we develop this strategy while focusing on the changes in vertex engagement. Here we don't want it decrease after adding a new vertex. Thus we have: if $n\notin \mathcal{AS}$, $\mathbb{D}_{u,\mathcal{AS}}$ $\geq$ $k$, and $\frac{d_{u,\mathcal{AS}}+w_{C_{\mathcal{H}}}(u,x)}{\sum_{v\in \mathcal{AS}}d_{v,\mathcal{AS}}+w_{C_{\mathcal{H}}}(m,n)+ac(m)}>\frac{d_{u,\mathcal{AS}}}{\sum_{v\in \mathcal{AS}}d_{v,\mathcal{AS}}}$, we execute $Q$.push($n$) and $\mathcal{AS}$.push($n$). 

Since this expanding strategy is also a greedy algorithm, we set the $\mathcal{AS}$ to be a fixed one (which exists on line 15 before dealing the vertices in $K$) in practical calculations.

\stitle{\textbf{Occurrence-driven Strategy.}}
In this strategy, we try to let the alternative subgraph $\mathcal{AS}$ stop further expanding once it form a connected $k$-core. That's to say we need make sure that $\mathcal{AS}$ generates towards the subgraph where $u$ has the highest engagement level. Considering that how the engagement level of $u$ in temporal subgraph $\mathcal{AS}$ is influenced when adding a new vertex. Let $\mathcal{AS}^{'}$ be the new subgraph after a new vertex is involved, we can easily have that $\Delta d_u=d_{u,\mathcal{AS}^{'}}$ - $d_{u,\mathcal{AS}}$ and $\Delta d_{\mathcal{AS}\rightarrow \mathcal{AS}^{'}}$=$\sum_{v\in {\mathcal{H}^{'}}}$ $d_{v,\mathcal{AS}^{'}}$ - $\sum_{v\in \mathcal{H}}d_{v,\mathcal{H}}$. When we have $\Delta E$=$ Eng_u(\mathcal{AS}^{'})$-$ Eng_u(\mathcal{AS})$, we can get that $\Delta E$=$\frac{d_{u,\mathcal{AS}^{'}}} {{\sum_{v\in \mathcal{AS}^{'}}}d_{v,\mathcal{AS}^{'}}}$ - $\frac{d_{u,\mathcal{AS}}}{\sum_{v\in \mathcal{AS}}d_{v,\mathcal{AS}}}$=
$\frac{d_{u,\mathcal{AS}}+\Delta d_u}{\sum_{v\in \mathcal{AS}}d_{v,\mathcal{AS}}+\Delta d_{\mathcal{AS}\rightarrow \mathcal{AS}^{'}}}$ - $\frac{d_{u,\mathcal{AS}}}{\sum_{v\in \mathcal{AS}}d_{v,\mathcal{AS}}}$=$(\frac{\Delta d_u}{\Delta d_{\mathcal{AS}\rightarrow \mathcal{AS}^{'}}}- Eng_u(\mathcal{AS}))\frac{\Delta d_{\mathcal{AS}\rightarrow \mathcal{AS}^{'}}}{\sum_{v\in \mathcal{AS}^{'}}d_{v,\mathcal{AS}^{'}}}$. If we want the $\Delta E$ not to decrease at each round, so that we can get a good result as soon as possible, we need to maximize $\Delta d_u$ and minimize $\Delta d_{\mathcal{AS}\rightarrow \mathcal{AS}^{'}}$. Since $Eng_u(\mathcal{AS})$ is a constant value at each round, we can just ignore it. A feasible greedy solution for this is that we add all $u$'s neighbors which meet the requirement for degree into $\mathcal{AS}$, and we just need to make $\Delta d_{\mathcal{AS}\rightarrow \mathcal{AS}^{'}}$ as small as possible. To improve the simplicity of operation, we just consider the value of $o_{(m,n)}(min(\mathcal{T_H}),max(\mathcal{T_H}))$ instead of the possible value of $d_{n,\mathcal{AS}}$. That's means we want $o_{(m,n)}(min(\mathcal{T_H}),max(\mathcal{T_H}))$ has the minimum value, that's why we call this occurrence-driven strategy. The strategy is that: if $n\notin AS$, $\mathbb{D}_{n,AS}$ $\geq$ $k$, and
    $o_{(m,n)}(min(\mathcal{T_H}),max(\mathcal{T_H}))$ is minimum, we execute $Q$.push($n$) and $AS$.push($n$). Until every vertex $x$ in $\mathcal{AS}$ that $\mathbb{D}_{x,AS}$ $\geq$ $k$.

Here we formally introduce our local search algorithm \textit{BULS}. When dealing with the first cumulative graph $C_{\mathcal{H}}$, we use the algorithm \ref{alg: naive candidate} to generate the alternative graph. Besides, for the cumulative graphs in following steps, we use the algorithm \ref{alg:advanced} with reference-strategy to expand, the rest of process is the same with \textit{TDGP}. Since the algorithm \ref{alg: naive candidate} lacks efficiency, we develop another two enhanced local search algorithms \textit{BULS+} and \textit{BULS*}. Specifically, for \textit{BULS+} we use the algorithm \ref{alg:advanced} with engagement-driven strategy to deal the first cumulative graph $C_{\mathcal{H}}$. The difference between \textit{BULS+} and \textit{BULS*} is that \textit{BULS*} replaces the engagement-driven strategy with the occurrence-driven strategy. Additionally, since these two more effective algorithms might miss the results in some cases, we will turn to use the algorithm \ref{alg: naive candidate} when we find there is no such $k$-core containing $u$ after the expanding using them for the alternative subgraph. 

\subsection{Analysis}
In this algorithm, since we use a BFS way to traverse the cumulative graph, the time complexity for each round of expanding is $O(n)$. It should be noticed that since this local search process will be executed in every accumulate graphs, the time complexity of the whole algorithm is still $O(nm^{2})$. As for the space complexity, it is unchanged as $O(nm)$. 

\section{Experimental Evaluation}
\label{sec:experiments}
Several comprehensive experiments are carried out to assess efficiency, effectiveness, and scalability of the proposed solutions. These experiments are executed on a server with an Intel Xeon 2.50GHZ CPU and 32GB memory running Ubuntu 18.04.
\subsection{Experimental setup}
\stitle{Datasets description.} We evaluate our solutions on real-world temporal networks \footnote{
http://snap.stanford.edu/, http://konect.cc/, http://www.sociopatterns.org/} with different types and sizes (Table \ref{tab:data}), including social (Facebook, Twitter, Wiki), email (Enron, Lkml), and scientific collaboration (DBLP) networks.  It should be noticed that we ignore the self-loops and treat all the graphs as undirected graphs.

\begin{table}[!t]
\begin{center}
\begin{minipage}{260pt}
\caption{Datasets statistics.}\label{tab:data}%
\begin{tabular}{@{}llllll@{}}
\toprule
Dataset & $n=\mid V \mid$ & $m=\mid\mathcal{E}\mid$ & $\bar{m}=\mid E \mid$  & $\mid\mathcal{T}\mid$ &   TS\footnotemark[1]\\
\midrule
 Facebook & 45,813 & 461,179 & 183,412 & 223 & Week\\
			 Twitter & 304,198 & 464,653 & 452,202 & 7 & Day \\
			 			 Wiki & 1,094,018 & 3,224,054 & 2,787,967 & 77 & Month\\
 Enron & 86,978 & 697,956 & 297,456 & 177 & Week \\
			 			 Lkml & 26,885 & 328,092 & 159,996 & 98 & Month\\
 DBLP & 1,729,816 & 12,007,380 & 8,546,306 & 72 & Year\\
\botrule
\end{tabular}
\footnotetext[1]{TS is the time scale of the timestamp.}
\end{minipage}
\end{center}
\end{table}

\stitle{Algorithms.} To solve our significant engagement community search problem, we develop four algorithms: \textit{TDGP}, \textit{BULS}, \textit{BULS+}, and \textit{BULS*}. Additionally, to evaluate the efficiency and effectiveness of our proposed solutions, we take the \textit{TopkDBSOL} \cite{DBLP:journals/pvldb/ChuZYWP19} and \textit{CST}
\cite{DBLP:conf/sigmod/CuiXWW14} as baseline models for  comparison. \textit{TopKDBSOL} is the online algorithm to find the top-k density bursting subgraphs, here the value of k is 100. \textit{CST} refers to the algorithm to handle the problem of community search with threshold constraint. For \textit{SECS} and \textit{CST}, the default value of parameter $k$ is 2, that every vertex has a degree no less than $k$. For \textit{TopKDBSOL}, the default value of parameter $\theta$ is 3, that the minimum time period of density bursting community is 3. To be more reliable, we randomly select 100 vertices as query vertices and report the average running time and quality. All algorithms are implemented in C++. 

\begin{table}[h]
\begin{center}
\begin{minipage}{260pt}
\caption{Running time of different algorithms with default parameters (second)}\label{tab:data1}%
\begin{tabular}{@{}lllllll@{}}
\toprule
 & DBLP & Lkml & Enron  & Facebook &  Twitter & Wiki\\
\midrule
 \textit{CST} &109.04 &1.07&2.22 &1.55 &3.74 &30.71\\
	        \textit{TopkDBSOL}&1,707& 2,178& 1,920 &39  &26,872 &20,217\\
 \textit{TDGP} & 279.38  &25.49  & 71.77 &70.60 &5.45 & 30.35\\ 
		     \textit{BULS} & 111.90 &5.36  & 12.98 &18.42 &6.28 & 25.55 \\
			 \textit{BULS+} & 36.27  &5.04  & 8.55 &10.49 &4.36  & 16.48\\
			 \textit{BULS*} & 20.15  &2.09 & 1.81 &2.18 &2.79 &5.15\\ 
\botrule
\end{tabular}

\end{minipage}
\end{center}
\end{table}

\begin{figure}[!t]
    \begin{minipage}{0.3\linewidth}
    \centerline{\includegraphics[width=4.0cm]{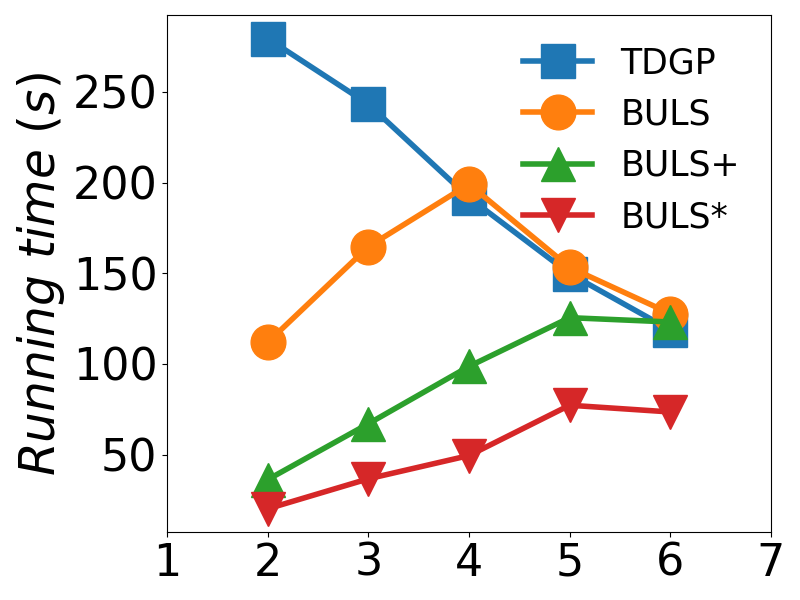}}
    \centering
    \caption{Running time of different strategies with various $k$ in Dblp}\label{fig:exp2:a}
    \end{minipage}
    \hfill
    \begin{minipage}{0.3\linewidth}
    \centerline{\includegraphics[width=4.0cm]{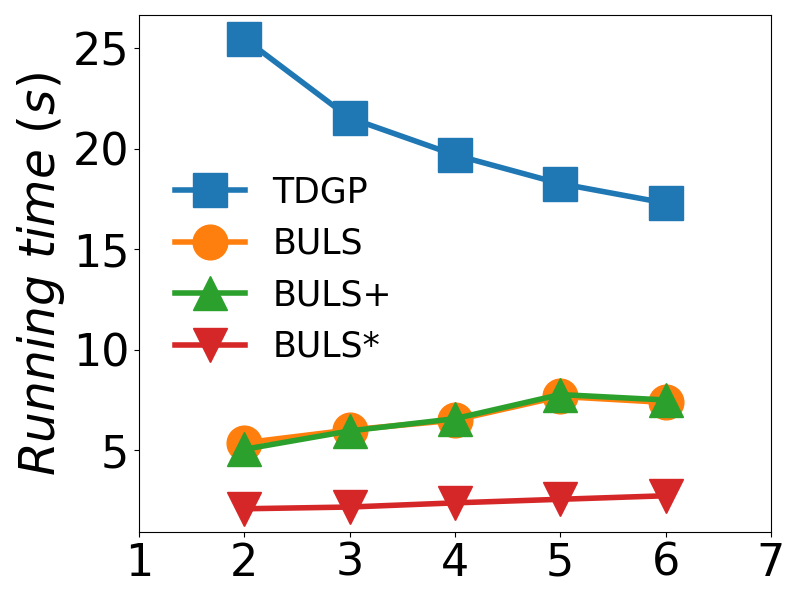}}
    \caption{Running time of different strategies with various $k$ in Lkml}\label{fig:exp2:b}
    \end{minipage}
    \hfill
    \begin{minipage}{0.3\linewidth}
    \centerline{\includegraphics[width=4.0cm]{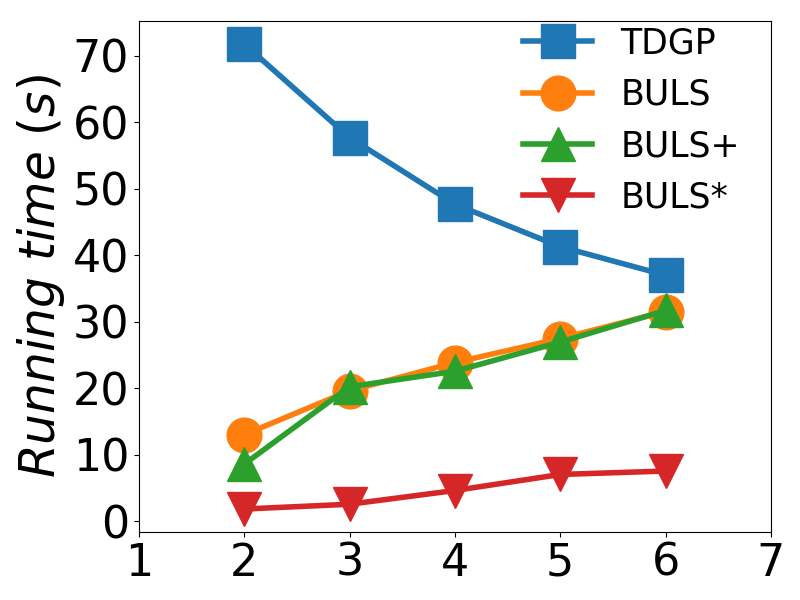}}
    \caption{Running time of different strategies with various $k$ in Enorn}\label{fig:exp2:c}
    \end{minipage}
\end{figure}

\subsection{Efficiency testing} 
\stitle{Exp-1: Running time of different algorithms with default parameters.} Table \ref{tab:data1} shows the  running time of different algorithms with default parameters.  For our  problem, the \textit{BULS*} has the least running time, which proofs that our \textit{BULS*} has excellent performances. Besides, \textit{BULS*} is even faster than \textit{CST} in DBLP, Enron, Twitter and Wiki, because we just consider the part around the query vertices instead of the whole graph.

\stitle{Exp-2: Running time of different strategies with various $k$.} To test how the parameter $k$ affect the running time of \textit{TDGP}, \textit{BULS}, \textit{BULS+} and \textit{BULS*}, we vary  $k$ from 2 to 6 (Fig. \ref{fig:exp2:a}-\ref{fig:exp2:c}). The \textit{BULS*} has the best performance in most cases.The running time of \textit{TDGP} decrease with increasing $k$, this is because when $k$ is bigger, the size of $k$-core will be smaller, which relieve the task for following steps. However, for \textit{BULS}, \textit{BULS+} and \textit{BULS*}, the running time has an upward trend at the beginning with the rise of $k$, the reason is that when the $k$ is bigger, the subgraphs that meet the constraints for $k$-core usually have a bigger scale, then the $bestresult$ will be smaller. As a result, the alternative graph have a bigger size. We can also observe that with further increasing of $k$, in Fig. \ref{fig:exp2:a} the running time of $\textit{BULS}$, $\textit{BULS+}$ and $\textit{BULS*}$ slightly descend due to the same reason mentioned for the $\textit{TDGP}$. That the total amount of vertices that need to be dealt with decreases.

\begin{figure}[!t]
    \begin{minipage}{0.48\linewidth}
    \centerline{\includegraphics[width=4.0cm]{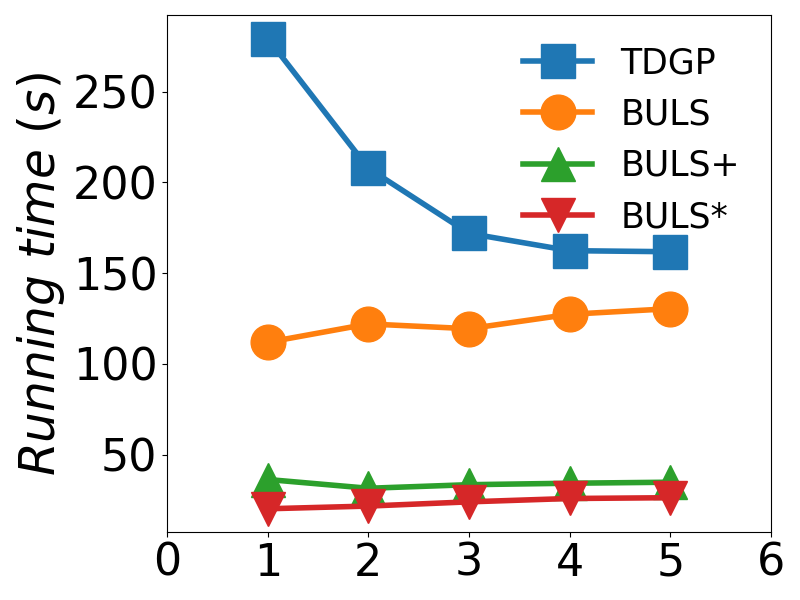}}
    \centering
    \caption{Scalability testing (VTS)}\label{fig:sct:a}
    \end{minipage}
    \hfill
    \begin{minipage}{0.48\linewidth}
    \centerline{\includegraphics[width=4.0cm]{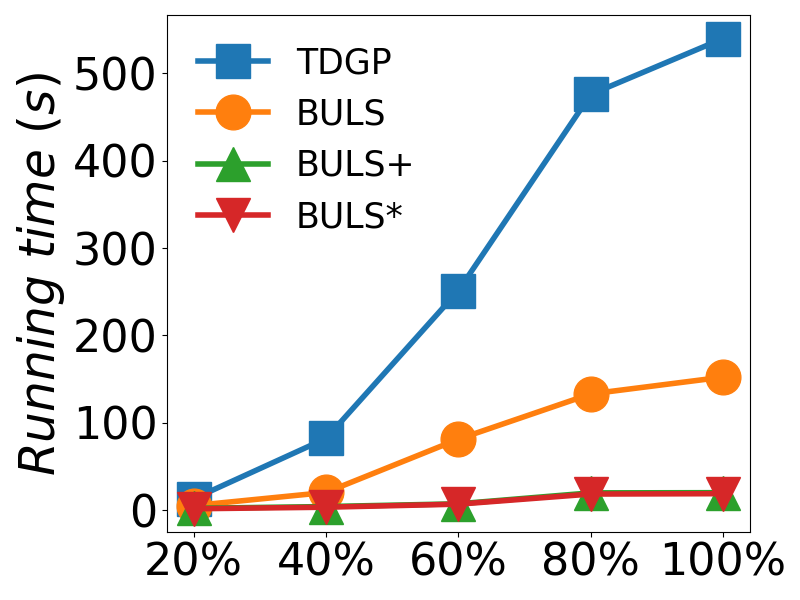}}
    \caption{Scalability testing (VNS)}\label{fig:sct:b}
    \end{minipage}

\end{figure}

\stitle{Exp-3: Scalability testing.}
To test the scalability of our algorithms. We generate additional temporal graphs from DBLP in two various ways. Specifically, the first one sets the sampling time scale from 2 years to 5 years and get the corresponding graphs, obviously, the temporal graphs tend to be denser with larger time scale. Another method is that we randomly select $20\%$,$40\%$,$60\%$,$80\%$ amount of nodes from the original graph and induce the new temporal subgraphs. We can observe the results in Fig. \ref{fig:sct:a} and \ref{fig:sct:b}, where VTS and VNS respresent varying time scale and varying nodes sampling. Clearly, the \textit{BULS*} has the least running time both in two experimental setups. The overall state in Fig. \ref{fig:sct:a} is similar to those in Fig. \ref{fig:exp2:a} generalized to similar explanations about the structural properties. Besides, the result in Fig. \ref{fig:sct:b} reveals that our local search strategy has the low global sensitivity attribute.   

\subsection{Effectiveness testing}  

Here, we use three metrics to evaluate the quality of the results. Specifically, they are engagement level ($EL$), temporal density ($TD$) \cite{DBLP:journals/pvldb/ChuZYWP19} and temporal conductance ($TC$) \cite{DBLP:conf/www/SilvaSS18}. The engagement level has been defined in Definition \ref{def:eng}, which focus on the activeness of the query vertex within the community. Temporal density is the metric that measures the denseness of the community. Formally, for a community $\mathcal{S}$, its temporal density $TD(S)$ is defined as $2*\mid(u,v,t)\in \mathcal{E}_{S}\mid/\mid S \mid(\mid S\mid-1)\mid\mathcal{T}_{S}\mid$, where $\mid S\mid$ is the amount of vertices in $S$, $TD$ reveals the structure density of the community. As for the temporal conductance, $TC(S)=Tcut(S, V\backslash S)/min(Tv(S),Tv(V\backslash S))$, where $Tcut(S,V/S)=\mid\{(u,v,t)\in\mathcal{E}\mid u\in S, v\in V/S\}\mid$ and $Tv(S)=\mid\{ (u,v,t)\in \mathcal{E}\mid u\in S, v\in S\}\mid$. the temporal conductance reveals the difference on the structural density between the inner part and outside part of community, with smaller $TC$, a community tends to be tighter separating from the original temporal graph.  

\stitle{Exp-4 :Effectiveness of different temporal methods with default parameters.} the results presented in Table \ref{tab:data2}-\ref{tab:data4} show the effectiveness of different community search methods in several temporal graphs. For the $EL$, we can observe that the \textit{BULS*} algorithm can get the communities with the maximum $EL$ in every dataset, because we can get the alternative graphs which are more query-centered and finally lead to the phenomenon that local search algorithms receive better results. Similarly, the results in $TD$ reveal that \textit{BULS}, \textit{BULS+} and \textit{BULS*} all can get more dense communities comparing to the \textit{TDGP}. As for the $TC$ metric, \textit{BULS}, \textit{BULS+} and \textit{BULS*} get the result no good than \textit{TDGP}, because they need to further delete the vertices in the subgraphs which have already been quite dense. 

\begin{table}[!t]
\begin{center}
\begin{minipage}{300pt}
\caption{Enagement Level ($EL$) of different temporal methods with default parameters }\label{tab:data2}%
\begin{tabular}{@{}lllllll@{}}
\toprule
 Method & DBLP & Lkml & Enron  & Facebook &  Twitter & Wiki\\
\midrule
			  \textit{TopkDBSOL} & 0.08066 & 0.11402 &0.11594 & 0.30513 & 0.27436 &0.16327\\
 \textit{TDGP} & 0.24863  &0.08374  & 0.04616 & 0.07378 & 0.02021 & 0.00347\\ 
		      \textit{BULS} & 0.26550  &0.16695  & 0.19599 & 0.19171 & 0.02213 & 0.02213 \\
			  \textit{BULS+} & 0.33252  &0.26043  & 0.25149 & 0.23799 & 0.19746  & 0.12075\\
			 \textit{BULS*} & 0.33296 & 0.26126 & 0.27353 & 0.25269 & 0.20566 & 0.12542\\

\botrule
\end{tabular}

\end{minipage}
\end{center}
\end{table}

\begin{table}[!t]
\begin{center}
\begin{minipage}{300pt}
\caption{Temporal Density ($TD$) of different temporal methods with default parameters }\label{tab:data3}%
\begin{tabular}{@{}lllllll@{}}
\toprule
 Method & DBLP & Lkml & Enron  & Facebook &  Twitter & Wiki\\
\midrule
			 \textit{TopkDBSOL} &0.62719 &0.86461 &0.87161 &0.72440
			 &0.54550 &0.72971\\
 \textit{TDGP} & 0.49613  &0.12067  & 0.03411 &0.03024 &0.04541 & 0.00114\\ 
		      \textit{BULS} & 0.50345  &0.14237  & 0.08595 &0.06515 &0.04861 & 0.00729 \\
			 \textit{BULS+} & 0.61016  &0.29831  & 0.12180 &0.06577 &0.44427  & 0.12443\\ 
			  \textit{BULS*} & 0.60115 & 0.31070 & 0.13437 & 0.07473 & 0.47565 & 0.13657\\

\botrule
\end{tabular}

\end{minipage}
\end{center}
\end{table}

\begin{table}[t!]
\begin{center}
\begin{minipage}{300pt}
\caption{Temporal Conductance ($TC$) of different temporal methods with default parameters }\label{tab:data4}%
\begin{tabular}{@{}lllllll@{}}
\toprule
 Method & DBLP & Lkml & Enron  & Facebook &  Twitter & Wiki\\
\midrule
\textit{TopkDBSOL} &0.71291 &0.96631 &0.93400 &0.60309
			 &0.97360 &0.98143\\
\textit{TDGP} & 0.64446  & 0.68174 & 0.80379 &0.75149 &0.69343 &0.82614 \\ 
		     \textit{BULS} & 0.65812  & 0.74965 &0.90476 &0.79464 &0.69495 &0.82809 \\
			  \textit{BULS+} & 0.78363  & 0.82302 & 0.93326 &0.08033 &0.84034  &0.87694\\ 
			  \textit{BULS*} & 0.78911 & 0.84447 & 0.97193 & 0.83826 & 0.93164 & 0.96230\\
			 
\botrule
\end{tabular}

\end{minipage}
\end{center}
\end{table}

\begin{figure}[!t]
    \begin{minipage}{0.3\linewidth}
    \centerline{\includegraphics[width=4.0cm]{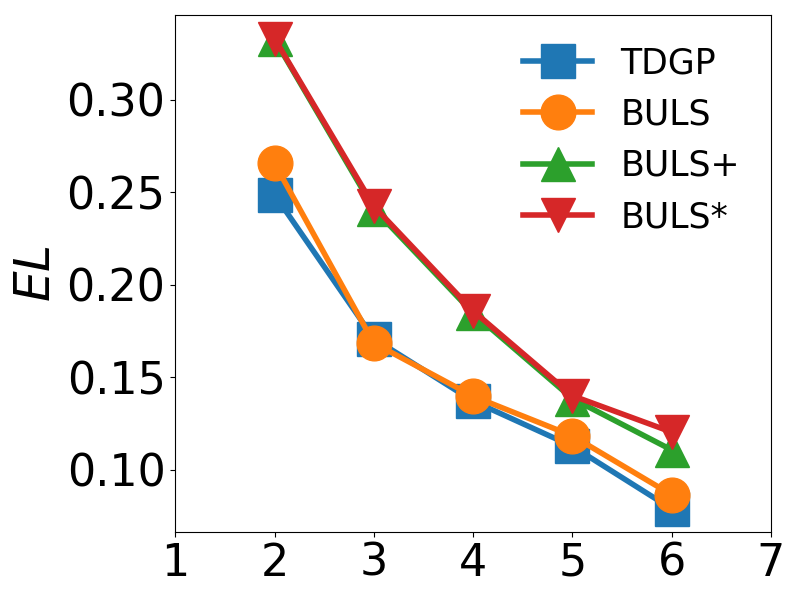}}
    \centering
    \caption{$EL$ comparison with different $k$ value in Dblp}\label{fig:subfig3:a}
    \end{minipage}
    \hfill
    \begin{minipage}{0.3\linewidth}
    \centerline{\includegraphics[width=4.0cm]{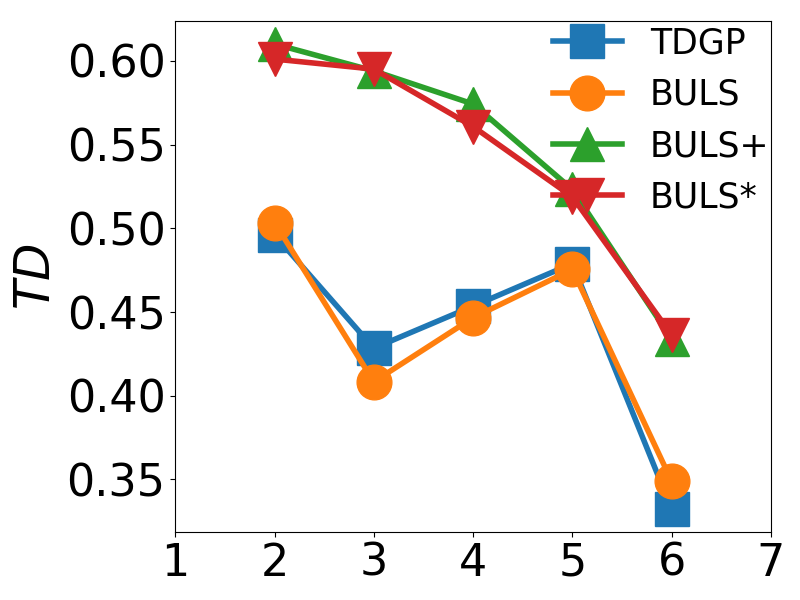}}
    \caption{$TD$ comparison with different $k$ value in Dblp}\label{fig:subfig3:b}
    \end{minipage}
    \hfill
    \begin{minipage}{0.3\linewidth}
    \centerline{\includegraphics[width=4.0cm]{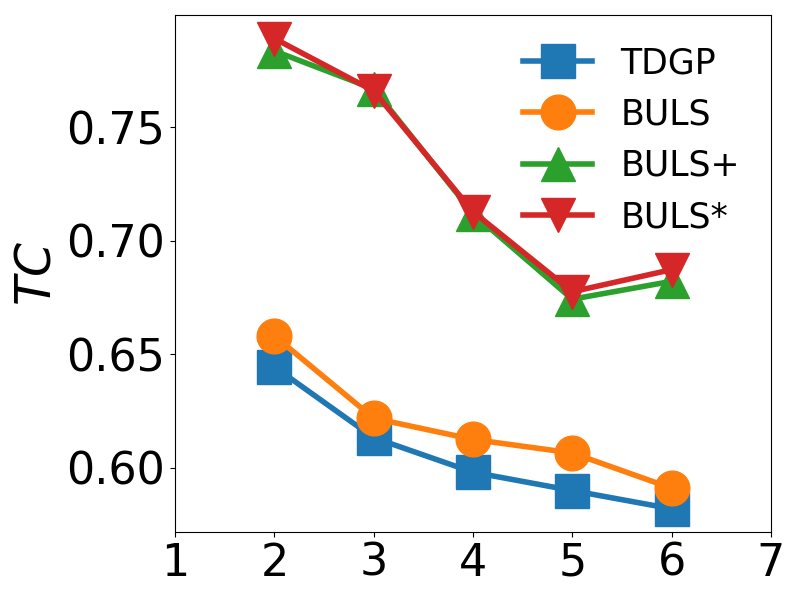}}
    \caption{$TC$ comparison with different $k$ value in Dblp}\label{fig:subfig3:c}
    \end{minipage}
        \begin{minipage}{0.3\linewidth}
    \centerline{\includegraphics[width=4.0cm]{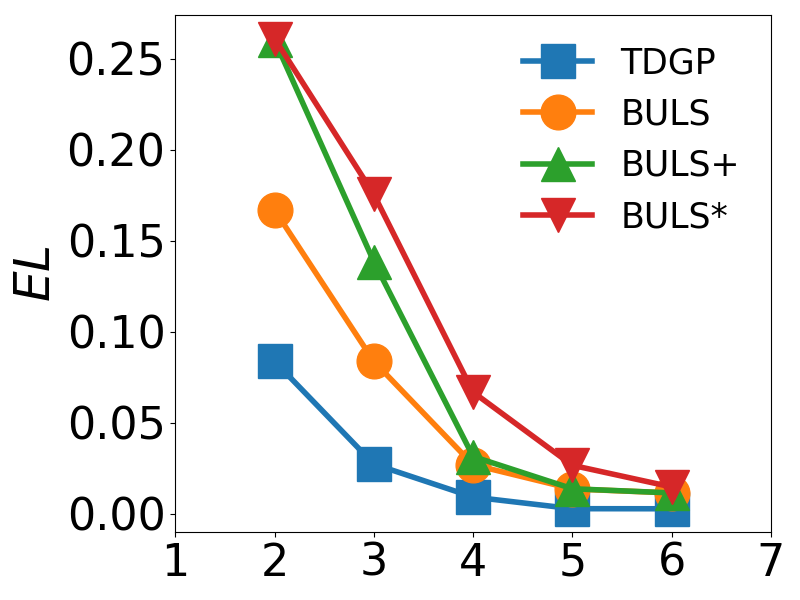}}
    \centering
    \caption{$EL$ comparison with different $k$ value in Lkml}\label{fig:subfig3:a2}
    \end{minipage}
    \hfill
    \begin{minipage}{0.3\linewidth}
    \centerline{\includegraphics[width=4.0cm]{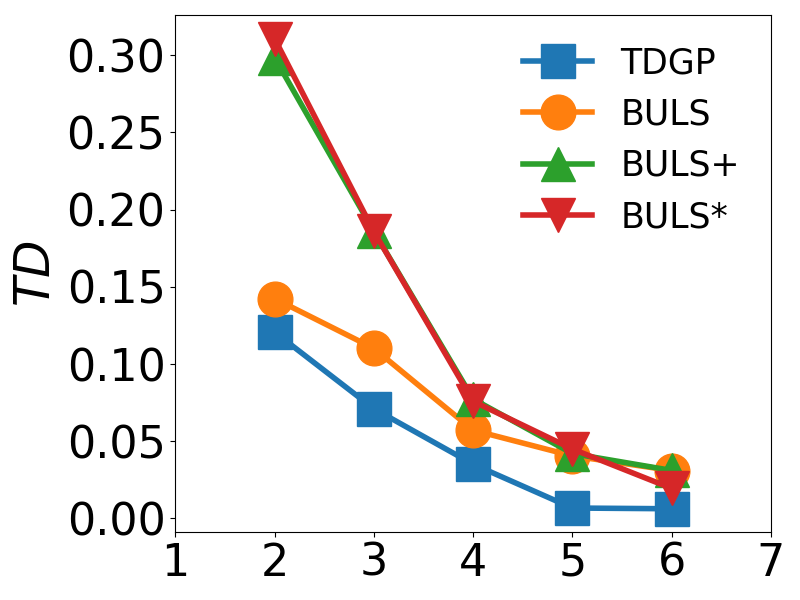}}
    \caption{$TD$ comparison with different $k$ value in Lkml}\label{fig:subfig3:b2}
    \end{minipage}
    \hfill
    \begin{minipage}{0.3\linewidth}
    \centerline{\includegraphics[width=4.0cm]{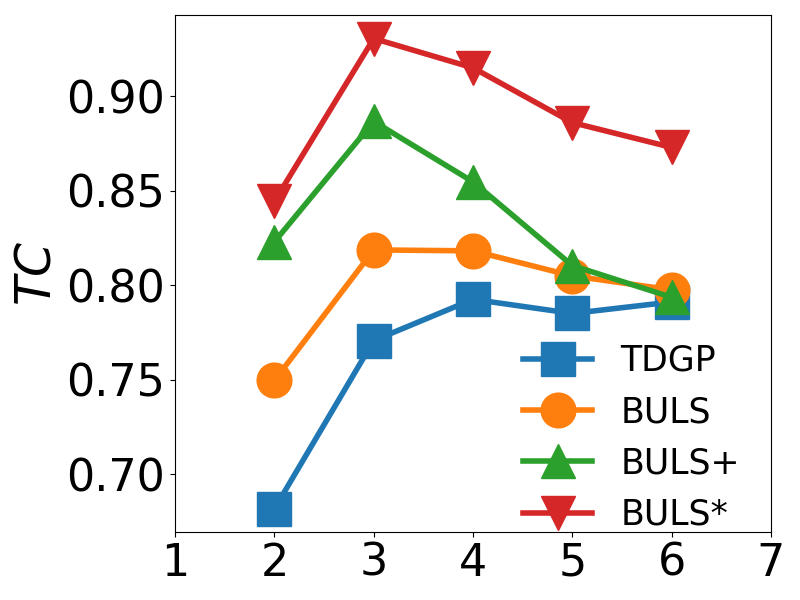}}
    \caption{$TC$ comparison with different $k$ value in Lkml}\label{fig:subfig3:c2}
    \end{minipage}
\end{figure}

\stitle{Exp-5: Effectiveness metrics comparison with different $k$ value.}
 Fig. \ref{fig:subfig3:a}-\ref{fig:subfig3:c2} demonstrate how the value of effectiveness metrics change with the value of $k$, we can observe that the relationships among the four algorithms are similar to those in Exp-5, which proof that our algorithms can still have good performance with the raising of $k$. Additionally, the values of $EL$ and $TD$ are reducing when $k$ grows bigger, for that the communities have more vertices. 
 
\begin{figure}[!t]
    \begin{minipage}{0.3\linewidth}
    \centerline{\includegraphics[width=4.0cm]{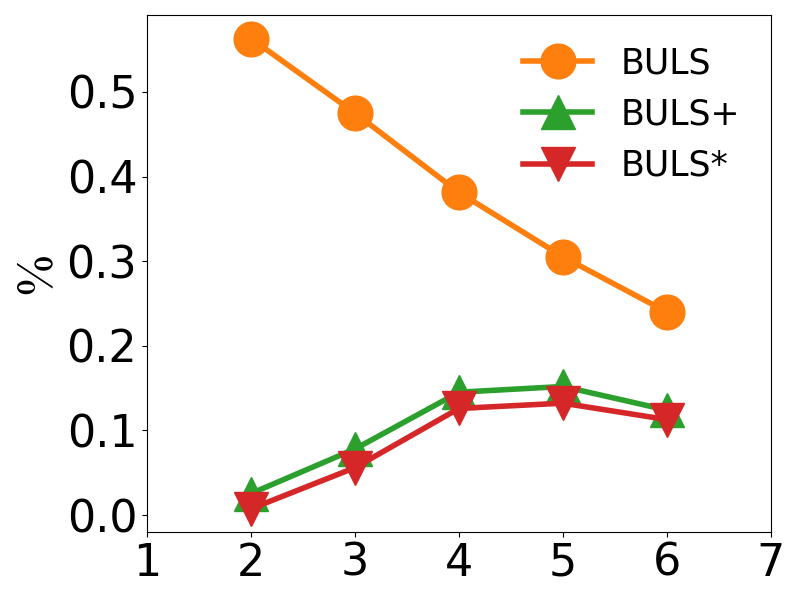}}
    \centering
    \caption{The size of expanded graphs of different strategies with various $k$ in Dblp}\label{fig:exp21:a}
    \end{minipage}
    \hfill
    \begin{minipage}{0.3\linewidth}
    \centerline{\includegraphics[width=4.0cm]{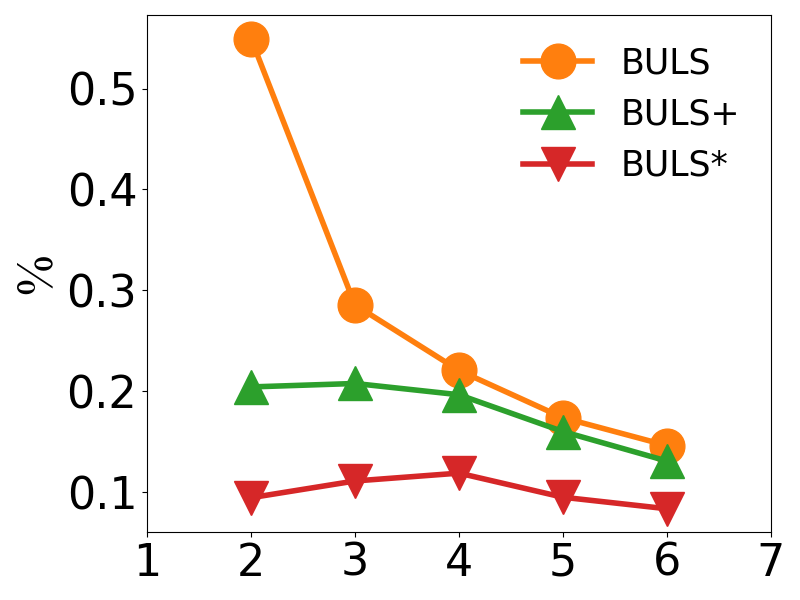}}
    \caption{The size of expanded graphs of different strategies with various $k$ in Lkml}\label{fig:exp21:b}
    \end{minipage}
    \hfill
    \begin{minipage}{0.3\linewidth}
    \centerline{\includegraphics[width=4.0cm]{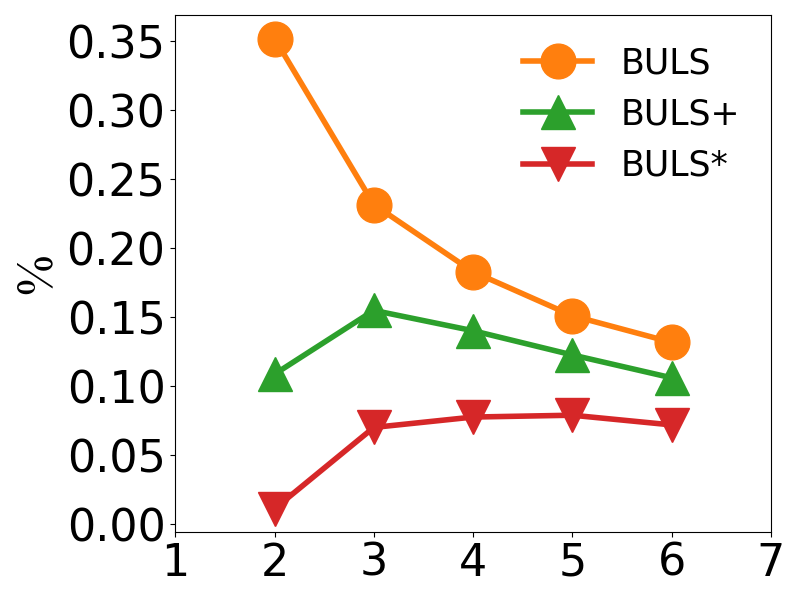}}
    \caption{The size of expanded graphs of different strategies with various $k$ in Enorn}\label{fig:exp21:c}
    \end{minipage}
\end{figure}

\stitle{Exp-6: The size of expanded graph in with various $k$.}
In order to observe how our local search algorithms perform from the aspect of reducing the search space, we count percentage of the size of the first expanded graphs of the original graphs, using different expanding strategies under different circumstances of $k$. The results are presented in Fig. \ref{fig:exp21:a}-\ref{fig:exp21:c}. We can see that with $\textit{BULS*}$ we can get the expanding graphs with the smallest size in all three temporal graphs, which makes it has the most efficient. The results using $\textit{BULS}$ decrease with the increasing of $k$, for that less vertices will be included since fewer of vertices have larger values of degree. However, the other two algorithms have a slightly increase at the beginning with the raising of $k$. For $\textit{BULS*}$, it takes more vertices to form a core with a larger $k$ value.

\begin{figure}[!t]
    \begin{minipage}{0.3\linewidth}
    \centerline{\includegraphics[width=4.0cm]{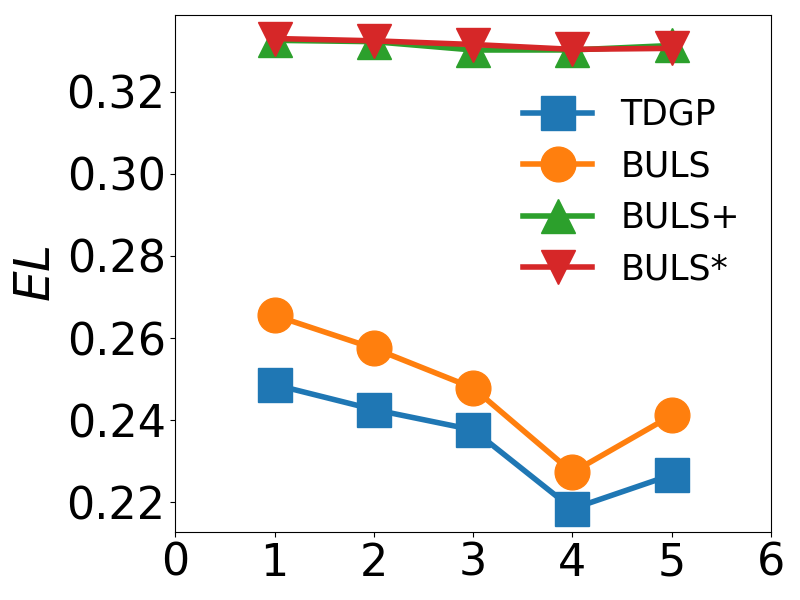}}
    \centering
    \caption{VTS ($EL$)}\label{fig:scef:a}
    \end{minipage}
    \hfill
    \begin{minipage}{0.3\linewidth}
    \centerline{\includegraphics[width=4.0cm]{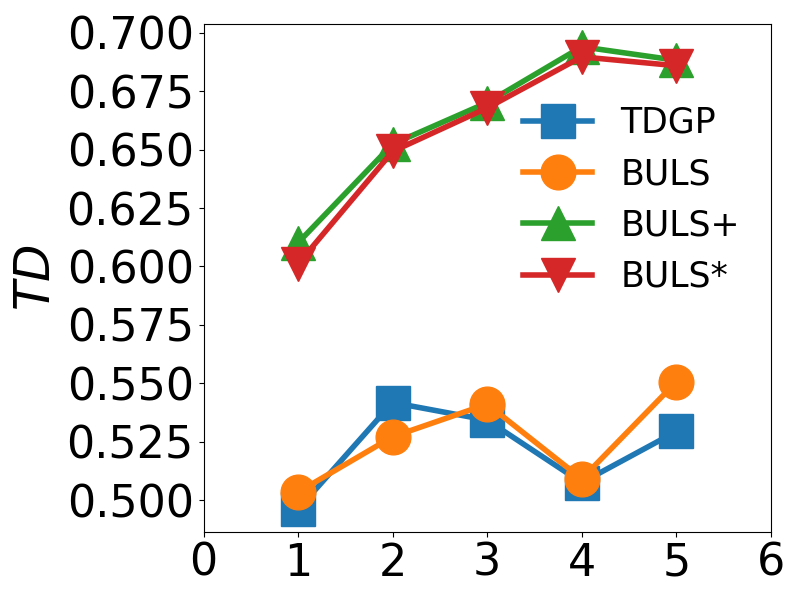}}
    \caption{VTS ($TD$)}\label{fig:scef:b}
    \end{minipage}
    \hfill
    \begin{minipage}{0.3\linewidth}
    \centerline{\includegraphics[width=4.0cm]{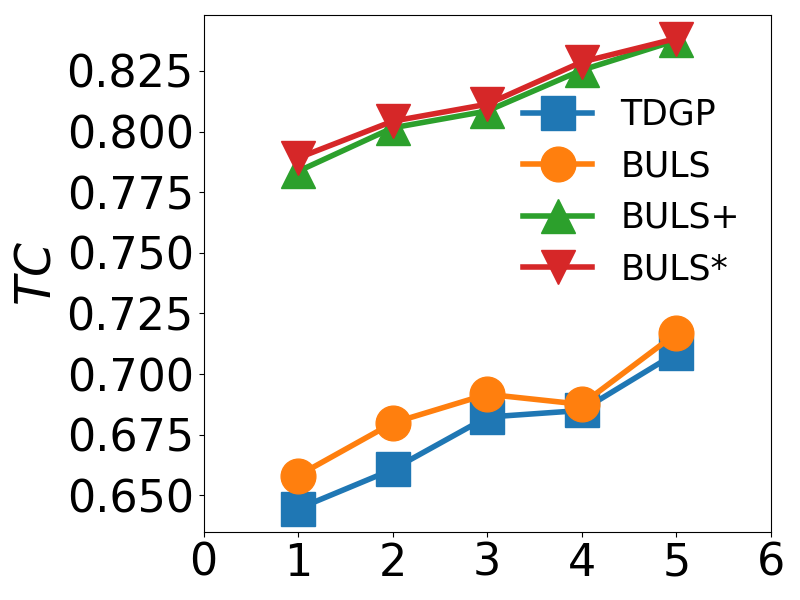}}
    \caption{VTS ($TC$)}\label{fig:scef:c}
    \end{minipage}
    \end{figure}
    \begin{figure}
        \begin{minipage}{0.3\linewidth}
    \centerline{\includegraphics[width=4.0cm]{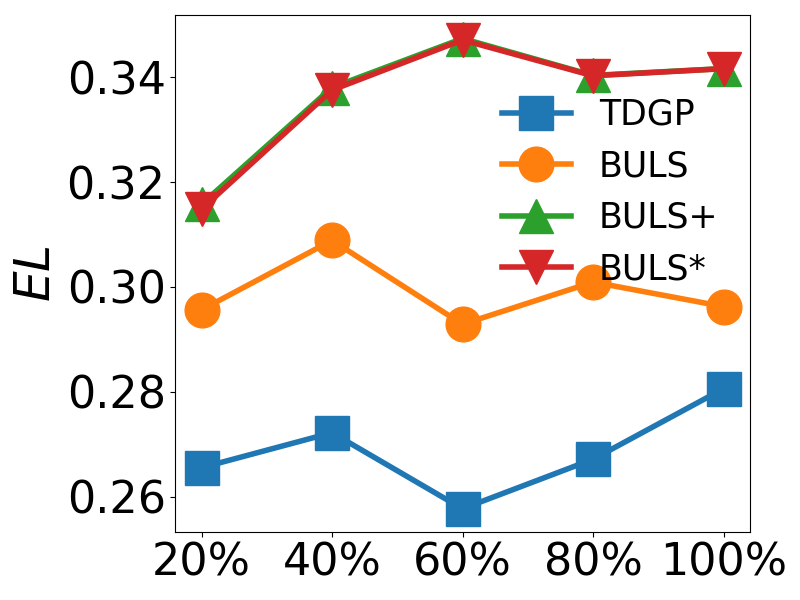}}
    \centering
    \caption{VNS ($EL$)}\label{fig:scef:d}
    \end{minipage}
    \hfill
    \begin{minipage}{0.3\linewidth}
    \centerline{\includegraphics[width=4.0cm]{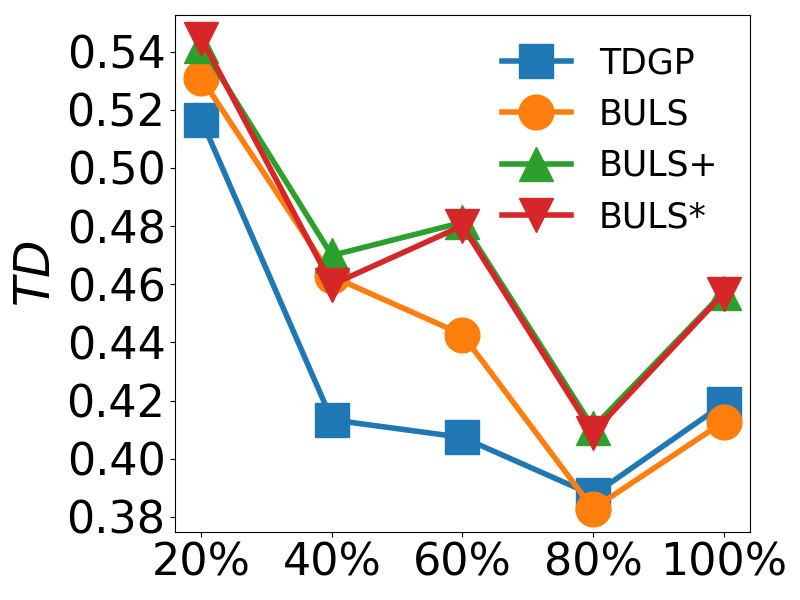}}
    \caption{VNS ($TD$)}\label{fig:scef:e}
    \end{minipage}
    \hfill
    \begin{minipage}{0.3\linewidth}
    \centerline{\includegraphics[width=4.0cm]{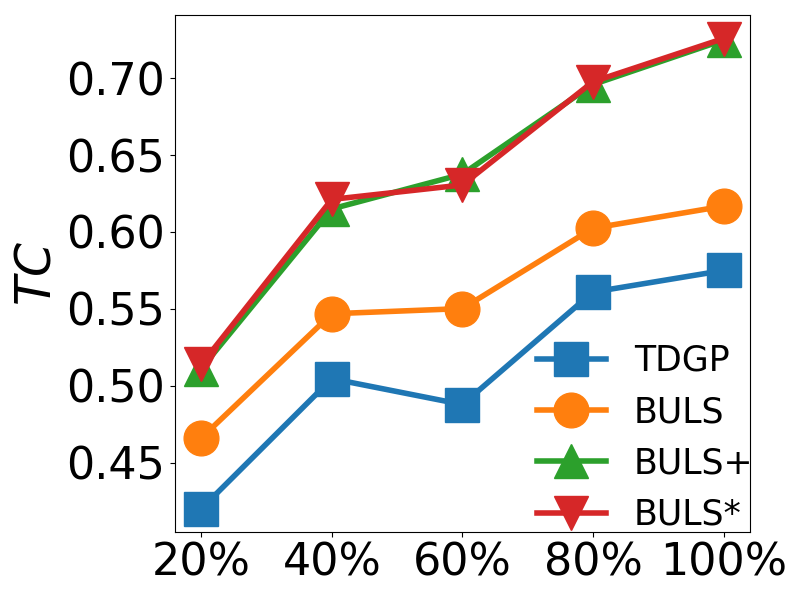}}
    \caption{VNS ($TC$)}\label{fig:scef:f}
    \end{minipage}
\end{figure}

\stitle{Exp-7: Scalability test for effectiveness metrics.}
Here we evaluate the scalability of our three algorithms with the effectiveness metrics. Similarly, we generate the subgrpahs from DBLP. Fig. \ref{fig:scef:a}-\ref{fig:scef:c} present the the results on subgraphs generated by VTS and Fig. \ref{fig:scef:d}-\ref{fig:scef:f} are those results running on the subgraphs generated by VNS. We can observe that $\textit{BULS+}$ and $\textit{BULS*}$ can get the similar good results under all the circumstances.

\begin{figure}[!t]
    \begin{minipage}{0.3\linewidth}
    \centerline{\includegraphics[width=4.0cm]{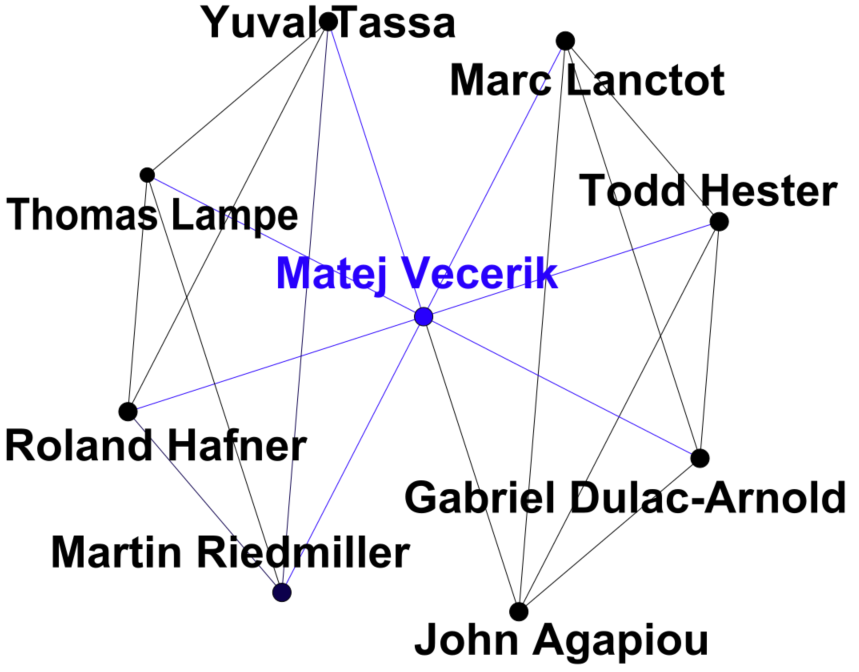}}
    \centering
    \caption{Matej (\textit{SECS})}\label{fig:cs:a}
    \end{minipage}
    \hfill
    \begin{minipage}{0.3\linewidth}
    \centerline{\includegraphics[width=4.0cm]{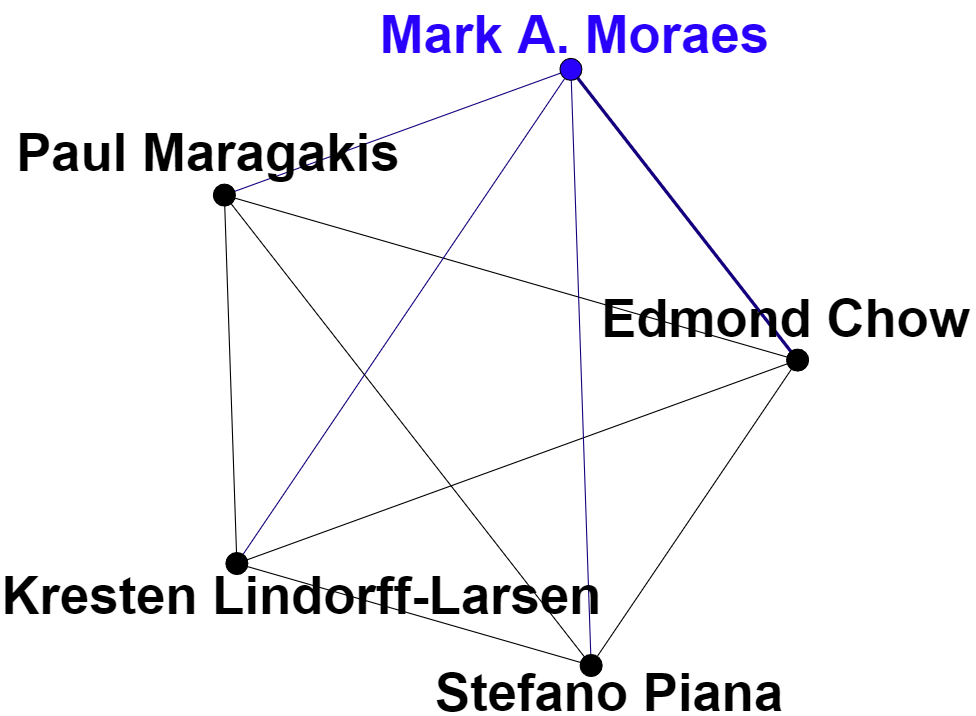}}
    \caption{Mark (\textit{SECS})}\label{fig:cs:b}
    \end{minipage}
    \hfill
    \begin{minipage}{0.3\linewidth}
    \centerline{\includegraphics[width=4.0cm]{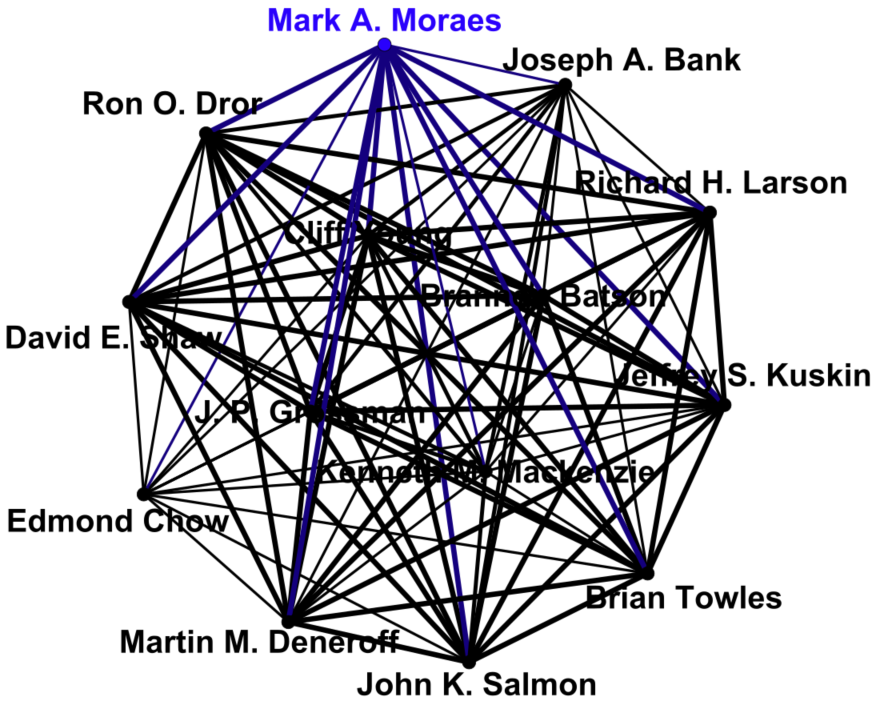}}
    \caption{Mark (\textit{TopkDBSOL})}\label{fig:cs:c}
    \end{minipage}

\end{figure}

\stitle{Exp-8: Case study on DBLP}
The Fig. \ref{fig:cs:a} and \ref{fig:cs:b} show the significant engagement community search  for the query vertex Matej Vecerik and Mark A. Moraes using our model. The Fig. \ref{fig:cs:c} is the result obtained by \textit{TopkDBSOL}. Note that the thicker the edges, the larger edge occurrences them has. We can observe that in the results of our model for the query vertices, the vertices tends to be the center roles of the communities, that have more temporal edges connect to other members, which reveal that the properties the communities carry are important for the person, in comparison in Figure \ref{fig:cs:c} Mark A. Moraes is not so impressive for the community is dense overall.

\section{Related Work}\label{sec:relate}
\subsection{Community mining}
Community is a general concept appears in physics, computational biology, and computer science, and so on \cite{Fortunato2009Community}. Notable methods include modularity optimization \cite{newman2004fast},  spectral analysis \cite{donetti2004detecting}, hierarchical clustering \cite{rokach2005clustering} and cohesive subgraph discovering \cite{DBLP:conf/icde/ChangQ19}. Typically, these methods are collectively known as community detection, which aims to identify all the communities from graphs, resulting in that are query-independent and time-consuming.  As a meaningful counterpart, community search has recently been proposed for semi-supervised learning task that can recover the community in which the query vertex is located \cite{DBLP:conf/icde/HuangLX17, DBLP:journals/vldb/FangHQZZCL20}. These methods mainly focus on searching user-specified communities on simple graphs or attributed graphs. For simple graphs, they aim to identify the communities that contain the given query vertices and satisfy a specific community model such as $k$-core \cite{DBLP:conf/kdd/SozioG10,DBLP:conf/sigmod/CuiXWW14, DBLP:journals/datamine/BarbieriBGG15}, $k$-truss \cite{DBLP:conf/sigmod/HuangCQTY14, DBLP:conf/sigmod/LiuZ0XG20}, clique \cite{DBLP:conf/sigmod/CuiXWLW13, DBLP:journals/tkde/YuanQZCY18}, density \cite{DBLP:journals/pvldb/WuJLZ15}, and connectivity \cite{DBLP:conf/kdd/TongF06, DBLP:conf/sigmod/RuchanskyBGGK15}. For instance, Sozio et al. \cite{DBLP:conf/kdd/SozioG10} introduced a framework of community search, which requires the target community is a connected subgraph containing query vertices and has a good score w.r.t. the proposed quality function. In particular, they used the  $k$-core as the quality function. Since the $k$-core is not necessarily dense, Huang et al. \cite{DBLP:conf/sigmod/HuangCQTY14} adopted a more cohesive subgraph model $k$-truss to model the community.  Recently, Wu et al. \cite{DBLP:journals/pvldb/WuJLZ15} observed the above approaches exist free-rider effect, that is, the return community often contains many irrelevant vertices to the query vertices. They proposed query-biased density to force the densest subgraph that is near the query vertices. Besides simple graphs, more complicated attribute information associated with vertices and edges also has been investigated for community search. Such as community search on keyword-based graphs \cite{DBLP:journals/pvldb/FangCLH16, DBLP:journals/pvldb/HuangL17, DBLP:conf/icde/LiuZZHXG20}, location-based social networks \cite{DBLP:journals/pvldb/FangCLLH17, DBLP:conf/kdd/ChenL0XY020}, multi-valued graphs \cite{DBLP:conf/sigmod/LiQYYXXZ18}, and heterogeneous information networks \cite{DBLP:journals/pvldb/FangYZLC20, DBLP:journals/pvldb/JianWC20}. However, they ignore the temporal properties of networks that frequently appear in applications. Thus it is unclear how to apply these techniques to solve our problem.

\subsection{Temporal networks mining.}
Temporal networks as a powerful paradigm that can model the complex networks in a fine-grained manner, in which each interaction between vertices occurs at a specific time. As a result, many problems and algorithms on temporal networks have been investigated \cite{holme2012temporal,DBLP:journals/corr/Holme15a}. For example,  Huang et al. \cite{DBLP:conf/sigmod/HuangFL15} considered the problem of minimum spanning tree (\textit{MST}), and they modeled two temporal \textit{MST} based on time and cost, where the term \textit{time} regarded as earliest arrival times and the term \textit{cost} treated as smallest total weight. In \cite{DBLP:conf/edbt/ZufleREF18,  DBLP:journals/tkde/SemertzidisP19}, these two works researched the temporal pattern matching problem.   Kumar et al. \cite{DBLP:journals/pvldb/KumarC18} investigated the temporal motif called temporal cycles, and they pointed out the temporal cycle motif can distinguish different temporal graphs by doing qualitative experiments.  In order to store and analyze efficiently the massive temporal graphs data, Wu et al. \cite{DBLP:conf/dasfaa/WuZCY17} proposed a \textit{equal-weight damped time window model} that considers the important of data follow an exponential decay function, and the model aggregates the massive temporal graphs into $\theta$ weighted graphs, the parameter $\theta$ is used to balance between the computational efficiency and the loss cost. Until recently, some work have been done on community mining over massive temporal networks \cite{DBLP:conf/bigdataconf/WuCLKHYW15, DBLP:conf/kdd/YangYWCZL16, DBLP:conf/cikm/GalimbertiBBCG18, DBLP:conf/icde/LiSQYD18}. For example, Ma et al. \cite{DBLP:conf/icde/MaHWLH17} researched the densest temporal community problem in special temporal graphs, in which nodes and edges are unchanged while the weights of edges are change over time. The densest temporal community is modeled as a set of vertex, which have the maximum the sum of weight. Yang et al. \cite{DBLP:conf/kdd/YangYWCZL16} proposed the concept of $\gamma$-dense in the temporal graph.  the $\gamma$-dense demands the result is a $\gamma$-quasi-clique at any timestamp of a given interval. Core decomposition problem is investigated on temporal networks in \cite{DBLP:conf/cikm/GalimbertiBBCG18, DBLP:conf/bigdataconf/WuCLKHYW15}. Lin et al. \cite{9351686} investigated the diversified lasting cohesive subgraphs on temporal graphs. Qin et al. \cite{9210070} studied the periodicity of subgraphs and proposed the concept of periodic clique to character and predict periodicity of cohesive subgraphs.

\section{Conclusion}
\label{def:con}
In this paper, we first introduce the definition of engagement level, and then raise a novel problem called significant engagement community search. To tackle this problem, we develop a global algorithm called \textit{TDGP}. To further improve the efficiency, we then devise a local search algorithm called \textit{BULS} and its enhanced version \textit{BULS+} and \textit{BULS*}. Finally, we evaluate our solutions on six real-world temporal graphs and the results show the superiority of our solutions.

\textbf{Acknowledgements.} The conference paper \cite{DBLP:conf/dasfaa/ZhangLYJ22} which contains part content of this  research is first published in [LNCS 13245, 250-258, 2022] by Springer Nature. The research is supported by the National Key Research and Development Program of China (No. 2018YFB1402802), NSFC (Nos. 62072205 and 61932004).

\bibliography{my-reference}

\end{document}